\documentstyle[11pt,newpasp,twoside]{article}
\def\edcomment#1{\iffalse\marginpar{\raggedright\sl#1\/}\else\relax\fi}
\reversemarginpar

\def\spose#1{\hbox to 0pt{#1\hss}}

\begin{document}
\title{The Significance of the Sodium Detection in the Extrasolar 
Planet HD~209458~b Atmosphere}

\author{S.\ Seager}
\affil{Institute for Advanced Study, School of Natural Sciences, 1
Einstein Drive, Princeton, NJ 08540}

\affil{The Carnegie Institution of Washington, Dept. of
Terrestrial Magnetism, 5241 Broad Branch Rd. NW, Washington, DC
20015}

\begin{abstract}
The Hubble Space Telescope (HST) detection of an extrasolar planet
atmosphere in 2001 was a landmark step forward for the
characterization of extrasolar planets. HST detected the trace
element sodium, via the neutral atomic resonance doublet at 593
nm, in the transiting extrasolar giant planet HD~209458~b. In this
paper I discuss the significance of this first ever extrasolar
planet atmosphere detection. I explain how the sodium measurement
can be used as a constraint on HD~209458~b atmosphere models and
review recent interpretations of the lower-than-expected sodium
line strength.
\end{abstract}
\vspace{-4mm}
\section{Introduction}
The first---and only to date---atmosphere detection of an extrasolar
planet was made by Charbonneau et al. (2002) with the Hubble Space
Telescope (HST). Charbonneau et al. (2002) observed the parent star
HD~209458~A, during and outside of the planet transit. During a transit,
some of the stellar intensity passes through the optically thin parts
of the planet atmosphere. Thus the star's spectrum taken during
transit is expected to contain some weak signature of the planet
atmosphere. The planet atmosphere signature measured in this way is
called the ``transmission spectrum''.

The strength of the planetary transmission spectrum can be
estimated as the ratio of the planet atmosphere annulus area
to the star area. We can very roughly
estimate the maximum annulus area by assuming the atmosphere is 10
``pressure scale heights'' thick. The atmospheric pressure scale
height, $H$, is the characteristic vertical dimension in pressure:
the e-folding distance for pressure,
\begin{equation}
H = \frac{P}{g\rho} = \frac{kT}{mg},
\end{equation}
where $P$ is pressure, $g$ is surface gravity, $\rho$ is density, and
$m$ is atomic or molecular mass. With appropriate numbers for
HD~209458~b ($T=1000$~K, $g=8.6$~m s$^{-2}$, $m$ the mass of H$_2$),
$H\sim$~500~km\footnote{The scale height is constant only for an
isothermal atmosphere and is otherwise depth dependent; i.e. it
changes throughout the atmosphere. For our estimate it is sufficient
to use an average number.}. Ten times the pressure scale height is
5000 km, which is 5\% of the planetary radius. Note that the large
scale height works to great advantage; the much cooler
planet Jupiter with a higher surface gravity has
a scale height of 24 km; ten times this is only 0.3\% percent of
Jupiter's radius.

The above estimate for atmosphere size can be used to estimate the
maximum strength of the planet atmosphere transmission spectrum in the
combined star + planet spectrum. This maximum estimate uses the
assumption that at some wavelengths the gas is strongly absorbing
(i.e., opaque) out to 10 pressure scale heights.  Comparing the area
of the atmosphere annulus to the stellar disk, ignoring stellar limb
darkening, we get $1.5\times10^{-3}$.

Because the planet transmission spectrum is so weak, the planet
transmission spectrum must be searched for in the residuals of the
in-transit minus out-of-transit stellar observations. Furthermore,
the weakness of the planet signature requires that observers are
unable to take a broad spectrum to search for atomic and molecular
absorption features, but instead must concentrate observations on
a pre-chosen narrow wavelength region in order to get a high
enough signal-to-noise. Charbonneau et al. (2002) describe a very
careful analysis resulting in a measurement of the sodium (Na)
line strength of $2.3 \pm 0.57 \times$ 10$^{-4}$. Even so, the Na
detection is a 4-$\sigma$ result and follow-up confirming
measurements of sodium other absorption features (such as the
resonance doublet of potassium (K) (767.0~nm), and near-IR H$_2$O,
CO, and possibly CH$_4$) would be reassuring. Na---expected to be
the strongest absorber---found to be less than the maximum
estimate implies that the other species may have much smaller
signatures than expected.

The planet atmospheric transmission spectrum is wavelength-dependent;
at wavelengths where no absorbers are present the stellar intensity
will pass through the atmosphere unimpeded, whereas at (possibly)
neighboring wavelengths a strong absorber will allow no stellar
intensity to be transmitted. This wavelength dependency can be thought
of a planet radius being different sizes at different
wavelengths. Comparing the continuum to an absorption line was used to
make the Na line measurement.

\section{The Significance of the Sodium Detection}
The first ever atmosphere detection is a landmark in extrasolar
resesarch because atmosphere studies open a whole new window to
extrasolar planet characterization. Three main reasons why the
sodium detection in the atmosphere of HD~209458~b is so important
are described in the following subsections.

\subsection{CEGPs Are What we Expect to First Order}
First and foremost the detection of neutral Na confirms the very
basic postulate that CEGPs have atmospheres expected for their
equilibrium effective temperatures ($T_{eq}$). This is because
HD~209458~b is a representative of the class of the close-in
extrasolar giant planets (CEGPs) with semi-major axes $<0.05$~AU.
Physically, $T_{eq}$ is is the effective temperature
attained by an isothermal planet after it has reached complete
equilibrium with its star.
\begin{equation}
T_{eq} = T_* \left( R_*/2a\right)^{1/2}[f(1-A_B)]^{1/4},
\end{equation}
where $T_*$ is the stellar temperature, $R_*$ is the stellar
radius, $a$ is the semi-major axis, $A_B$ is the Bond albedo, and
$f$ is a parameter to describe the heat redistribution from the
day side where $f=1$ if the heat is evenly distributed or $f=2$ if
only the day side reradiates the energy. With the unknown Bond
albedo the CEGPs are expected to range in temperature from
approximately 1000 to 1500~K; Na is expected to exist in
neutral gaseous form in atmospheres for any $T_{eq}$ in this
range. In other words, for HD~209458~b, for a given distance from
the parent star, star temperature, and solar composition the
HD~209458~b atmosphere is to first order what is expected.

Even though at 2 x 10$^{-6}$ solar abundance Na is essentially a trace
element, the neutral Na spectral signature is extremely strong at
optical wavelengths due to a very strong resonance line. In addition
there are expected to be no other strong optical absorption lines;
most atoms are locked into molecules which tend to have transitions at
either UV wavelengths (electronic transitions) or at IR wavelengths
(rotational-vibrational transitions). In fact since cool T dwarfs with
similar effective temperatures to HD~209458~b have extremely deep and
broad Na and K resonance lines it is not a surprise that neutral Na
was measurable in the HD~209458~b atmosphere. See Figure~2 in Liebert
et al. (2000) for the first observational confirmation of Na and K as
the major optical absorbers in T dwarfs, from a spectrum of the T
dwarf SDSS1624.

\subsection{The Sodium Line was Predicted in Advance}
The presence of neutral Na in the atmosphere of HD~209458~b was first
predicted for transmission spectra by Seager \& Sasselov
(2000). Before the atmosphere detection the Na line doublet was also
studied by Brown (2001) and Hubbard et al.  (2001). The HST atmosphere
detection of the transiting planet is reassuring because it involved a
specialized observation that required advance knowledge of what
atmospheric feature to look for, as described in Sec. 1. The success
of the Na detection shows that applications of atmospheric physics to
planets in new environments can be successful and that model results
as guidelines to experimental design are reliable.

\subsection{The Sodium Line Strength is Weaker than Expected} The
measured value of the Na absorption line feature was lower than
predicted. This is not surprising because the models did not include
accurate treatment of secondary effects and inputs, almost all of
which should reduce the strength of the Na line. To illuminate the
significance of the low value of detected Na I used simple inputs to
my atmosphere model and computed the change in transit depth in
adjacent bands as was done with the real data in Charbonneau et
al. (2002; see Figure~1 in this reference).  The simple inputs
include: solar composition, cloud-free, and that the heat from the
incident stellar intensity is instantaneously redistributed around the
planet. The ``secondary effects'' are all omitted: no photoionization,
no photochemistry, no atmospheric circulation.  For this reason I
emphasize that the model is just one out of a large range of parameter
space, but that it is sufficient for the illustrative purposes here.

The computed transit depth in the Na line from the simple model is
$\sim9 \times 10^{-4}$, 12 sigma away from the observational
measurement.  This simple model is therefore completely ruled out by
the data.  In this simple, homogeneous model the abundance of Na would
have to be approximately 1000 times less than solar to match the
observed line strength. In other words, the low measured value of the
Na line is significant.

\section{Using the Sodium Observation as a Constraint on Model
Atmospheres}
\subsection{Transmission Spectra Models}
The transmission spectrum from the transiting extrasolar planet
HD~209458~b is straightforward to compute. Stellar light rays travel
along the line of sight through the planet atmosphere towards the
observer. These rays from the star suffer exponential attenuation
due to wavelength-dependent absorption from the planetary
atmosphere\footnote{Refraction in the planet's atmosphere is
negligible (Hui \& Seager 2002).},
\begin{equation}
I_{\lambda} = I_{\lambda,0}\exp(-\tau_{\lambda}),
\end{equation}
where $I_{\lambda}$ is intensity and $I_{\lambda,0}$ is initial
intensity originating from the star. $\tau_{\lambda}$ is the
optical depth
\begin{equation}
\tau_{\lambda} = \int_0^L n(l) \sigma_{\lambda} dl,
\end{equation}
where $n(l)$ is the number density of a given species,
$\sigma_{\lambda}$ is the absorption cross section, and $L$ is the
path length through the planetary atmosphere. $\tau_{\lambda}$ can
include many different absorption species and $n(l)$ and
$\sigma_{\lambda}$ depend on $T$ and $P$.
For a resonance line transition where $n_1$ is
the number density of atoms with electrons in the ground state and
subscript 2 refers to the $n=2$ state,
\begin{equation}
n_1 \sigma_{\lambda} = n_1 \frac{h \nu_0}{4 \pi} B_{12} \phi(\nu),
\end{equation}
where $B_{12}$ is an Einstein probability and $\phi(\nu)$ is the
line profile.

While the transmission spectra, via exponential attenuation shown in
equation~(3), is straightforward to calculate, it relies on the
underlying radial atmosphere structure via $T$, and $n_1$ (which is
related to $P$ and to the radiation field $J$); see Figure~1. This
underlying radial atmosphere structure is difficult to calculate. The
radiative transfer methods used to self consistently compute the
atmospheric structure in proximity to the central star and results can
be found in a variety of sources (e.g., Seager 1999; Barman et
al. 2002; Burrows \& Sudarsky 2003; Sudarsky et al. 2003a, 2003b).
The atmospheric temperature-pressure structure is required to
determine: the species abundance (from chemical equilibrium) that
factors into the number density for the opacity in equation (4); and
the temperature required for absorption coefficients, e.g.  equation
(5). There are many uncertainties in these atmosphere models. Most
notably, none of them consider atmospheric circulation which is
crucial for the redistribution of stellar irradiation (Showman \&
Guillot 2002; Showman \& Guillot 2003; Cho et al. 2003) that
determines the atmospheric temperature-pressure structure. In addition
photochemistry is omitted as is non-equilibrium chemistry. Steps are
being made to advance the models and this is an ongoing process.

Different parts of a transmission spectrum absorption line are formed
at different parts in the atmosphere. The transmission spectral line
is created by removing photons from the stellar beam. This absorption
process can be thought of as additive over many layers of the
atmosphere (see Figure~1). Deep in the atmosphere, the number density
of absorbing atoms is high and at the line center all of the stellar
photons will be absorbed, i.e., the part of the line formed deep in
the atmosphere is saturated.  In addition, deep in the atmosphere
pressure broadening will be strong, causing absorption away from the
line center---the line wings. High in the atmosphere there is little
pressure broadening and hence little contribution to the line
wings. Because the number density of absorbers is much lower compared
to the deep atmosphere, the line core is not saturated and actually is
formed in the upper atmosphere. Although the Charbonneau et al. (2002)
measurement does not resolve the spectral line, some information can
be garnered about the gross line shape because the measurement is deeper
in the narrow bands than the medium bands, and is absent from the wide
bands,

\begin{figure}[h]
\plotfiddle{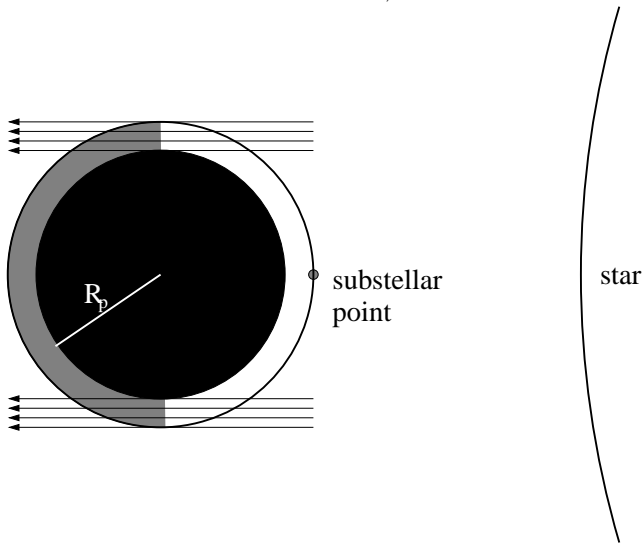}{2.5in}{0}{40}{40}{-130}{0} \caption{A
schematic cross section of HD~209458~b to illustrate transmission
spectrum modeling. Light rays (indicated by the lines with arrows)
travel from the star through the planetary atmosphere. The total
transmission spectrum is additive over the different rays. Because
the rays sample different radial atmosphere depths, the underlying
temperature-pressure structure is crucial to the transmission
spectrum calculation. This simple day/night picture for HD~209458~b,
however, is in error (Showman \& Guillot 2002, 2003; Cho et al.
2003). Note that for the transmission spectrum calculation
$R_p$---the measured planetary radius at optically thick
wavelengths---must be chosen.} \label{fig:2D}
\end{figure}

\subsection{Possible Interpretations of the Weak Sodium Line}

The Na line strength measured by Charbonneau et al. (2002)
provides the first constraint on extrasolar giant planet
atmosphere models. Notably the measured Na abundance is much lower
than predicted. This weaker than expected Na line strength can be
matched by only a few extreme cases in the parameter space modeled
by Charbonneau et al. (2002). In their discovery paper,
Charbonneau et al. (2002) outline four possible explanations for
the weaker than expected Na line: a high cloud; photoionization;
low Na abundance; or chemical equilibrum depletion of atomic Na
into molecules and solids.

Many causes have been proposed since the discovery paper, with
calculations to back them up, to explain the lower than expected
strength of the Na line.  The author finds it interesting that each
group finds the reason to be what their code can calculate. The
proposed explanations are described in the rest of this subsection.

{\bf Non-LTE effects} could make the Na line weaker and a different
shape compared to the simple model (Sec. 2.3).  LTE, an abbreviation
for ``local thermodynamic equilibrium'', is a simplification used in
most extrasolar planet atmosphere models (and many stellar atmosphere
models). The LTE assumption is that the state of the gas can be
described by only two variables, temperature and pressure. State of
the gas refers to the chemical equilibrium partitioning of molecules,
the ionization states, and, most relevant for this case, the atomic
level populations (used in equation (5)). In order to compute the
transmission spectrum of the Na line, the ground state level
population needs to be known (see equation~(4)). LTE is valid at high
densities and where the radiation field is a black body. In the upper
layers of HD~209458~b, where part of the transmission spectrum forms,
neither of these is valid and non-LTE is expected to prevail. The
atomic levels must be computed under non-LTE in order to compute the
line strength.  Because the upper atmosphere is most senstive to
non-LTE conditions, the non-LTE calculation mostly affects the line
core, even producing emission in the line core with absorption in line
wings. See Barman et al. (2002, 2003) for a more detailed description
of the non-LTE effect.

{\bf A high cloud} was first suggested in Seager \& Sasselov
(2000). We assumed that the cloud would be optically thick, and the
cloud top would correspond to $R_p$ in Figure~1. In this case the only
part of the planet atmosphere that is probed by the transmitting
stellar rays is the part of the atmosphere above the cloud tops. If
the cloud is high in the atmosphere, then the stellar rays probe a
small part of the atmosphere that is not very dense, with low pressure
broadening. The resulting Na transmission spectrum line is expected to
be fairly weak with no broad line wings. This is in contrast to the
low or no cloud case where $R_p$ is low in the atmosphere so that the
stellar rays probe a large fraction of the atmosphere where the high
number densities cause high pressure broadening. The resulting Na line
is expected to be strong and wide from pressure broadening.

{\bf Photoionization} is an obvious suggestion since due to the
proximity of the planet to the parent star the planet is receiving
huge amounts of stellar UV photons compared to solar system planets
where photoionization is known to play a role. Detailed
photoionization models have yet to be produced; in addition to Na they
would involve many atoms and molecules which could serve as sinks for
the UV radiation and sources of free electrons for Na$^+$ to
recombine. Using a simple photoionization model Fortney et al. (2003)
show that the weak Na line can be reproduced with photoionization
together with a high cloud.

{\bf Atmospheric circulation} models (Guillot \& Showman 2002; Cho et
al. 2003) are needed to describe the transfer of stellar radiation
throughout the planet atmosphere. HD~209458~b is expected to be
tidally locked (Guillot et al. 1996) to its parent star, thus one side
is permanently heated while the other side is in permanent
darkness. Strong winds are expected to redistribute the stellar
radiation, based on a comparison of the advective and radiative time
scales. Thus atmospheric circulation will determine the atmospheric
structure (see Figure~1) and hence the transmission spectrum. Guillot
\& Showman (2002, 2003) suggest that winds could transport atomic Na
from the planetary day side to the night side. They propose that
temperatures on the night side could be colder, allowing Na to
condense into solid NaCl where it could sink out of the atmosphere and
not be available as Na for transmission spectra. In this scenario, Na
is depeleted and a weak line results.

{\bf Post-accretionary extraplanetary origin of sodium} has been
suggested by Atreya et al. (2003).  Atreya et al. (2003) propose that
primordial Na is not present in HD~209458~b's atmosphere due to
differentiation to the planetary interior and subsequent depletion of
atomic Na into molecular species. They propose that the small amount
of Na present in HD~209458~b's atmosphere comes from later influx of
material from meteorite or comet impacts, planetary rings, or a
volcanically active satellite.

{\bf Low sodium abundance.} It could be difficult to attribute the
weakness of the Na line to a low Na abundance until all of the
above effects have been ruled out. As suggested in Charbonneau et
al. (2002), however, if observations of H$_2$O and CO show
``normal'' abundances compared to Na, then we may be able to
attribute the weak Na line to low primordial abundance.

{\bf Can anything be ruled out} from the Na measurement? There is no
detailed information about line shape from the Na measurement.
However, based on the fact that the transit in the Na line was weaker
in the medium band than the narrow band, and not detected in the wide
band, we can say that a very broad absorption line is not present.
Therefore we have an indication that either the deep atmosphere is not
being probed or the Na atomic number density is very low. Furthermore,
we know that photoionization alone is unlikely to be the only
explanation of the weak Na signal. Photoionization affects only the
upper atmosphere, and hence mostly the line core. In this case the
line should still be broad.  Emission line models (see Barman et
al. (2002)) under extreme assumptions in the non-LTE scenario are also
ruled out because a net increase---as opposed to a net decrease---in
Na line flux during transit would have been detected.
\vspace{-4mm}
\section{Other Observational Diagnostics}
The Na measurement is a very useful constraint on HD~209458~b's
atmosphere. Yet it is not enough to get anywhere near a clear picture
of the atmosphere. Many observations are ongoing to try to detect
other atmospheric constituents. We expect the H$_2$O near-IR
absorption bands to be prominent spectral features. CO or CH$_4$ or
both should also be present, the relative abundance is a good
temperature diagnostic (Seager, Whitney, and Sasselov 2000).
HD~209458~b's albedo and orbital light curve will be measured with the
upcoming space satellite MOST (Micro Oscillations of
STars)\footnote{http://www.astro.ubc.ca/MOST/2002/index.html}.
HD~209458~b's day and night side temperatures will be measured with
SIRTF. Following is a table of key features that are obtainable within
the next two years. The combination of all of these observational
diagnostics will provide powerful constraints on theoretical
atmosphere models.

\acknowledgements{This work was supported by the W. M. Keck
Foundation and by the Carnegie Institution of Washington.}

\begin{tabular}{l l l}
\hline \\
Bulk Property & Observational Diagnostic & Method \\
\hline
\hline
Temperature (T) &  CH$_4$ vs. CO  &   ground\\
\\
T of day & IR photometry & ground or SIRTF \\
and night side & & \\
\\
Albedo & Secondary eclipse photometry &  MOST \\
\\
General properties & Photometry &  MOST \\
of atmospheric scattering &during the planet's orbit \\
particles & & \\
\\
  Other spectral lines & Transit transmission spectra &
ground, HST \\
\\
\hline
\end{tabular}

\end{document}